\documentclass[10pt,conference]{IEEEtran}
\IEEEoverridecommandlockouts
\usepackage[noadjust]{cite}
\usepackage{amsmath,amssymb,amsfonts}
\usepackage{balance}
\usepackage{algorithmic}
\usepackage{graphicx}
\usepackage{textcomp}
\usepackage{xcolor}
\def\BibTeX{{\rm B\kern-.05em{\sc i\kern-.025em b}\kern-.08em
    T\kern-.1667em\lower.7ex\hbox{E}\kern-.125emX}}

\usepackage[capitalize]{cleveref}
\usepackage{xspace}
\usepackage{subcaption}
\usepackage{booktabs}
\usepackage{url}

\usepackage{glossaries}

\newacronym{csi}{CSI}{channel state information}
\newacronym{dl}{DL}{deep learning}
\newacronym{isac}{ISAC}{integrated sensing and communications}
\newacronym{iot}{IoT}{Internet of Things}
\newacronym{mimo}{MIMO}{multiple-input multiple-output}
\newacronym{ml}{ML}{machine learning}
\newacronym{mlp}{MLP}{multi-layer perceptron}
\newacronym{mse}{MSE}{mean squared error}
\newacronym{nn}{NN}{neural network}
\newacronym{ofdm}{OFDM}{orthogonal frequency-division multiplexing}
\newacronym{pca}{PCA}{principal component analysis}
\newacronym{plcp}{PLCP}{physical-layer convergence protocol}
\newacronym{sdr}{SDR}{software-defined radio}
\newacronym{vae}{VAE}{variational auto-encoder}

\newacronym{sq}{SQ}{scalar quantization}
\newacronym{vq}{VQ}{vector quantization}

\newcommand{\wifi}{\mbox{Wi-Fi}\xspace}

\begin{document}

\title{Efficient Wi-Fi Sensing for IoT Forensics\\ with Lossy Compression of CSI Data}

%\author{\IEEEauthorblockN{Anonymous Authors}}
\author{
    \IEEEauthorblockN{
        Paolo Cerutti, Fabio Palmese, Marco Cominelli, Alessandro E. C. Redondi
    }
    \IEEEauthorblockA{
        \textit{DEIB, Politecnico di Milano}\\
        Milan, Italy\\
    }
}

\maketitle

\begin{abstract}
Wi-Fi sensing is an emerging technology that uses channel state information (CSI) from ambient Wi-Fi signals to monitor human activity without the need for dedicated sensors.
Wi-Fi sensing does not only represent a pivotal technology in intelligent Internet of Things (IoT) systems, but it can also provide valuable insights in forensic investigations.
However, the high dimensionality of CSI data presents major challenges for storage, transmission, and processing in resource-constrained IoT environments.
In this paper, we investigate the impact of lossy compression on the accuracy of Wi-Fi sensing, evaluating both traditional techniques and a deep learning-based approach.
Our results reveal that simple, interpretable techniques based on principal component analysis can significantly reduce the CSI data volume while preserving classification performance, making them highly suitable for lightweight IoT forensic scenarios.
On the other hand, deep learning models exhibit higher potential in complex applications like activity recognition (achieving compression ratios up to 16000:1 with minimal impact on sensing performance) but require careful tuning and greater computational resources.
By considering two different sensing applications, this work demonstrates the feasibility of integrating lossy compression schemes into Wi-Fi sensing pipelines to make intelligent IoT systems more efficient and improve the storage requirements in forensic applications.
\end{abstract}


\section{Introduction}

\wifi sensing leverages existing wireless communication signals to monitor and interpret changes in the physical environment.
Specifically, the analysis of amplitude and phase variations in wireless signals enables innovative applications such as device-free motion detection, activity recognition, and environmental monitoring.
Unlike traditional sensing systems that rely on dedicated hardware, \wifi sensing builds upon the existing infrastructure of \wifi access points and stations, thereby reducing deployment costs and improving scalability for smart environments~\cite{jiang2018smart}.

Among the different approaches to \wifi sensing, \gls{csi} analysis is one of the most promising techniques~\cite{armentagarcia2024}.
The \gls{csi} provides an estimation of the wireless channel’s frequency response, measured for each frame at the \wifi receiver.
As the propagation of wireless signals is influenced by physical characteristics of the environment (such as room geometry, human motion, and the presence of objects), the \gls{csi} becomes a powerful proxy for monitoring environmental changes.

At the same time, the widespread adoption of \gls{iot} devices in smart environments has sparked interest in a new field of digital forensics, where network-derived data can be used as potential evidence~\cite{stoyanova}.
In this context, \gls{csi} emerges as a valuable source of information due to its correlation with human motion and environmental variations, and \wifi sensing can become a crucial tool for retrospective analysis, intrusion detection, anomaly identification, and other security applications.
For example, law enforcement may leverage historical \gls{csi} data to reconstruct movement patterns at crime scenes or detect unauthorized access in sensitive areas.

However, despite its advantages, \wifi sensing based on \gls{csi} can generate large volumes of data.
Modern IEEE 802.11ax transceivers can operate with signals on 160-MHz channels and 4x4 \gls{mimo} configuration, producing up to 32000 \gls{csi} data points for each received frame~\cite{gringoli2021axcsi}.
This poses significant challenges for real-time processing, storage, and transmission---especially in resource-constrained \gls{iot} scenarios.
On top of that, traditional forensic applications may require the data to be stored for very long periods of time before they are required for the investigation. 
As a consequence, optimizing the storage requirements is essential for improving the data collection and preservation pipeline.
While \emph{lossless} compression guarantees the full reconstruction of the original data, \emph{lossy} compression techniques can further improve the compression ratio at the cost of a (possibly small) reconstruction error, making \wifi sensing more efficient.

\begin{figure*}
    \centering
    \includegraphics[width=0.92\linewidth]{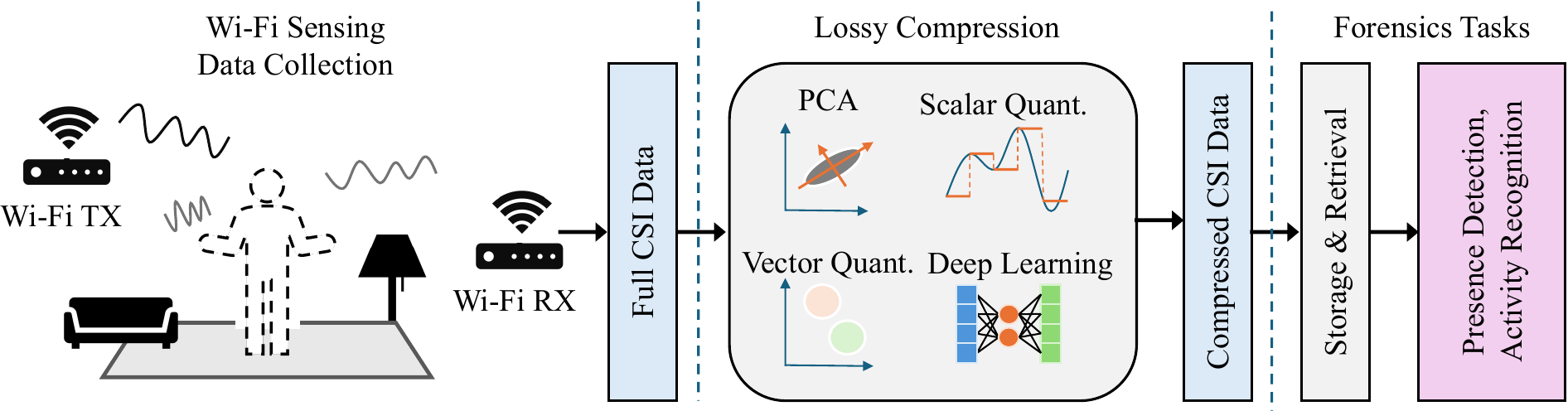}
    \caption{Overview of the workflow proposed for efficient \wifi sensing in \gls{iot} forensics applications. Collected \acrshort{csi} data is stored in a compressed form and retrieved after a potentially very long time.
    Lossy compression algorithms can be used to minimize the \acrshort{csi} storage requirements if they are proven to retain good sensing accuracy.}
    \label{fig:scenario}
\end{figure*}

In this work, we investigate the impact of lossy compression on \wifi sensing in forensic \gls{iot} contexts.
We evaluate both traditional compression schemes (using scalar and vector quantization) and \gls{dl} approaches based on recent works in this field~\cite{yang2022efficientfi, cominelli2023fusion}.
The pipeline we propose in \cref{fig:scenario} compresses \gls{csi} data at the time of collection, enabling efficient long-term storage and retrieval for forensic analysis.
Our goal is to characterize the tradeoff between compression ratio and classification accuracy for two important sensing tasks---presence detection and activity recognition---identifying strategies that preserve sensing capabilities while significantly reducing storage requirements.
To this end, we present and discuss empirical results obtained on real-world datasets showing that traditional compression techniques can already achieve outstanding performance with a simpler and more explainable architecture.
Our analysis ultimately aims to simplify the design of data-efficient and forensic-ready \wifi sensing systems.

\section{Background}
\label{sec:background}

% {\color{red}
% \subsection{IoT Forensics}
% \bf Preservation}

\newcommand{\mxn}{\ensuremath{\mathrm{M} \times \mathrm{N}}\xspace}
\newcommand{\bfy}{\mathrm{\mathbf{y}}}
\newcommand{\bfx}{\mathrm{\mathbf{x}}}
\newcommand{\bfH}{\mathrm{\mathbf{H}}}
\newcommand{\bfn}{\mathrm{\mathbf{n}}}
\newcommand{\numsc}{\mathrm{W}}

\subsection{CSI Extraction}
\label{ssec:csi-extraction}

In \wifi and in other \gls{ofdm} systems, the \gls{csi} is measured by estimating the physical properties of the communication channel, i.e., the attenuation and the phase delay of each \gls{ofdm} subcarrier.
The \gls{csi} is a critical element in wireless communications because it enables accurate equalization of frequency-selective distortion on wide-band wireless channels.

Let us consider the basic model for an \gls{ofdm} communication system with $\numsc$ subcarriers.
Each transmitted \gls{ofdm} symbol $\bfx \in \mathbb{C}^\numsc$ is distorted by the wireless channel $\bfH \in \mathbb{C}^\numsc$ and received as $\bfy \in \mathbb{C}^\numsc$ according to:
\begin{equation}
    \bfy = \bfH \circ \bfx + \bfn \, ,
    \label{eq:comm}
\end{equation}
where $\circ$ indicates the element-wise product and $\bfn \in \mathbb{C}^\numsc$ represents the channel noise.
Thanks to \cref{eq:comm}, the radio chipset at the receiver can directly estimate $\bfH$ by comparing the symbols received in the \wifi frames' preamble against some reference \gls{ofdm} symbols known as ``training fields''~\cite{khorov2019tutorial}.
In practice, given the specific signal in the training fields, the receiver measures the value of $\bfH$ for each subcarrier using techniques like Least Squares (LS) or Minimum Mean Squared Error (MMSE) on the expected and received sequences \cite{biguesh2006training}.
\Cref{eq:comm} assumes single-antenna systems, but it can be easily generalized also to \mxn \gls{mimo} systems with $\mathrm{M}$ transmit antennas and $\mathrm{N}$ receive antennas. In such cases, the receiver can estimate the channel matrix for each transmit-receive antenna pair, obtaining \mxn complex values per subcarrier for each frame.

% Depending on the specific modulation, different training fields are involved in the reconstruction of the CSI matrix for a specific frame. In legacy frames, the Legacy Long Training Field (L-LTF) is used as a reference for estimating the channel response: this value contains a predefined sequence for each \gls{ofdm} subcarrier and allows the receiver to estimate the \gls{csi} matrix by comparing the fields of the received signal with the expected sequence. However, different modulations involve different frame fields: for 802.11n High Throughput (HT) frames, the HT Long Training Field is used (HT-LTF), while 802.11ac Very High Throughput (VHT) frames involve the VHT-LTF.

Traditionally, \wifi chipsets used the \gls{csi} just to equalize the received frame and then discarded the information.
However, over the last decade, researchers have been developing \emph{\gls{csi} extractors}, i.e., tools that enable access to raw \gls{csi} data to monitor physical variations in the surrounding environment.
Indeed, advanced analysis of the \gls{csi} variations can reveal sensitive information about human activities (\cref{sec:related}).
Since the \gls{csi} is measured very close to the physical layer, its operation is strongly tied to the firmware and hardware of the specific radio chipset.
It follows that each \gls{csi} extractor can generally work with only a limited subset of \wifi chipsets (\cite{halperin2011tool,gringoli2021axcsi}) and that the \gls{csi} data encoding often differs between different extractors.
% In \cref{tab:csi-extraction-tools}, we briefly report a non-exhaustive list of tools that have been developed for \gls{csi} extraction.
In this paper, we consider two separate datasets collected using two different \gls{csi} extractors (\cref{sec:methodology}).
However, both tools report a sequence of $\numsc$ complex values for every \wifi frame received, corresponding to the attenuation and phase delay of each \gls{ofdm} subcarrier.

% \begin{table}
%   \centering
%   \caption{Non-exhaustive list of common \acrshort{csi} extraction tools available for commercial hardware.}
%   \begin{tabular}{lcr}
%     \toprule
%     CSI Extractor & Wi-Fi version & max CSI points \\
%     \midrule
%     802.11n CSI Tool~\cite{halperin2011tool} & 802.11n & 30 \\
%     Atheros CSI Tool~\cite{athcsi2015} & 802.11n & 56 \\
%     Nexmon CSI~\cite{nexmoncsi2019} & 802.11ac & 4096 \\
%     PicoScenes~\cite{picoscenes2021} & 802.11ax & 7968 \\
%     AX-CSI~\cite{gringoli2021axcsi} & 802.11ax & 32768 \\
%     \bottomrule
%   \end{tabular}
%   \label{tab:csi-extraction-tools}
% \end{table}

\subsection{Lossy Compression of CSI Data}
\label{ssec:lossy-compression}

% In this work, we consider several compression techniques to reduce the dimensionality of \gls{csi} data while preserving the essential characteristics needed for accurate human sensing.
The high-dimensional nature of raw \gls{csi} data can make direct processing expensive in terms of both computational resources and storage required to hold the data.
In this work, we explore systematically lossy compression techniques to reduce the dimensionality of \gls{csi} data while preserving the essential characteristics needed for accurate human sensing.
The techniques we explore include traditional techniques such as principal component analysis of the \gls{csi} data, scalar quantization, and vector quantization, as well as more advanced techniques based on deep learning.
We briefly discuss them in the following.

\paragraph{\Gls{pca}}
We utilize the \gls{pca} as a lossy compression technique by projecting the original \gls{csi} data into a lower-dimensional subspace.
More specifically, the \gls{pca} is used to identify the orthogonal basis vectors that capture most of the variance in the data.
Then, by discarding the components with lower variance contributions, we effectively reduce the \gls{csi} data dimension while retaining the most ``informative'' features.

\paragraph{\Gls{sq}}
We utilize the Lloyd-Max quantizer to minimize the \gls{mse} between the original \gls{csi} data and its quantized representation.
We control the trade-off between the compression ratio and the reconstruction accuracy by tuning the number of quantization levels.
While this is one of the simplest compression techniques, its impact on \gls{csi} data and \wifi sensing applications is still unexplored.

\paragraph{\Gls{vq}}
In this case, \gls{csi} feature vectors are grouped into clusters, with each cluster represented by a centroid point.
\Gls{vq} is similar to clustering algorithms such as \emph{k-means}, where similar data points are assigned to the same cluster.
In particular, we utilize \gls{vq} to compress the high-dimensional \gls{csi} data by replacing each vector with its closest centroid.
By controlling the number of centroids (i.e., clusters), we can again control the tradeoff between compression efficiency and reconstruction fidelity.
However, \gls{vq} has more potential to reduce storage and computational requirements significantly.

\paragraph{\Gls{vae}}
We also assess an alternative algorithm based on a popular \gls{dl} architecture that can be used to perform a data-driven compression of \gls{csi} data.
In classification tasks (with $K$ target classes in the set 
$\mathcal{C}_K$), \glspl{vae} are generative models that can be used to effectively estimate the class-conditional probability $p(x \,|\, \mathcal{C}_k)$ of a \gls{csi} feature vector $x$ (with $1 \leq k \leq K$).
Specifically, \glspl{vae} encode the full \gls{csi} feature vectors into a probabilistic representation within a very small latent space.
This representation is generally more convenient than the one provided by traditional auto-encoders because it is more robust to noise and can also provide a flexible characterization of the latent space~\cite{cominelli2023fusion}.
In particular, \gls{vae} architectures have been recently shown to achieve a good balance between compression ratio and sensing performance in \gls{iot} scenarios~\cite{yang2022efficientfi}.
In this work, we follow the approach outlined in \cite{cominelli2023fusion} to represent the \gls{csi} feature vectors with just four values, corresponding to the four parameters ($\mu_1,\mu_2, \sigma_1, \sigma_2$) of a bi-dimensional Gaussian distribution.
\section{Datasets and Methodology}
\label{sec:methodology}

\begin{figure}
    \centering
    \includegraphics[width=0.96\columnwidth]{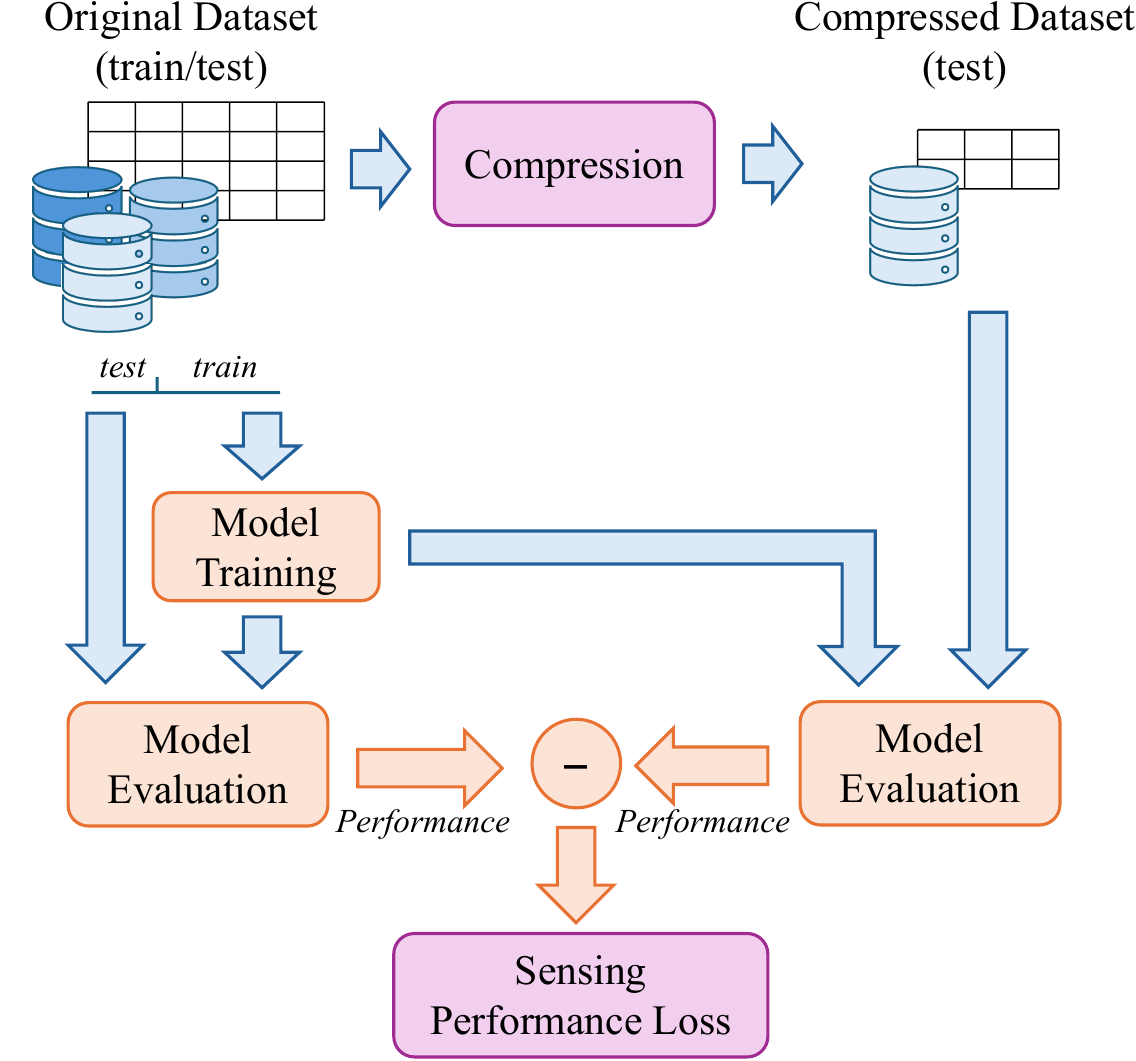}
    \caption{Methodology for evaluating the tradeoff between the compression ratio and sensing accuracy. In the compression stage, we consider several different algorithms; the sensing accuracy is evaluated in two target application scenarios, namely \textit{presence detection} and \textit{activity recognition}.}
    \label{fig:methodology}
\end{figure}

The diagram in \cref{fig:methodology} shows the processing pipeline used to analyze the tradeoff between the \gls{csi} data compression ratio and the sensing accuracy.
After data collection, the Compression block applies the lossy compression techniques described in \cref{ssec:lossy-compression}.
The uncompressed CSI data is split into two disjoint subsets for training and evaluating the models.
The trained models are then evaluated on the compressed data.
Instead of measuring the absolute sensing accuracy (which depends on the sensing model), we are more interested in the performance losses due to the data reconstruction errors from lossy compression.
Therefore, we measure the performance loss as the difference between the compressed vs. uncompressed sensing performance to analyze the tradeoff between storage and sensing capabilities.

\begin{figure}
    \centering
    \begin{subfigure}[b]{0.5\columnwidth}
        \includegraphics[width=\linewidth]{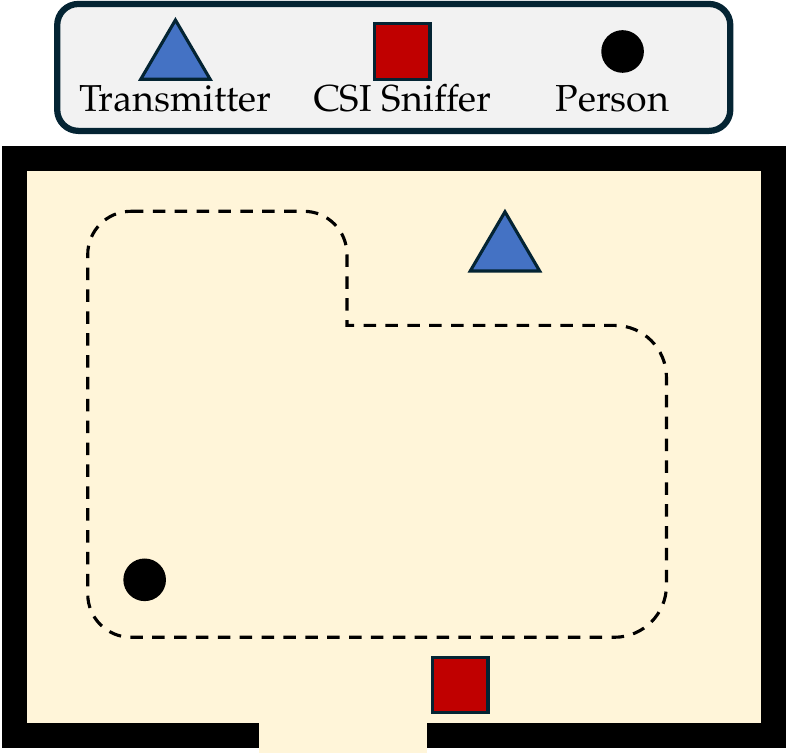}
        \caption{\textit{Presence detection} room.}
        \label{fig:lab1}
    \end{subfigure}
    \hfill
    \begin{subfigure}[b]{0.46\columnwidth}
        \includegraphics[width=\linewidth]{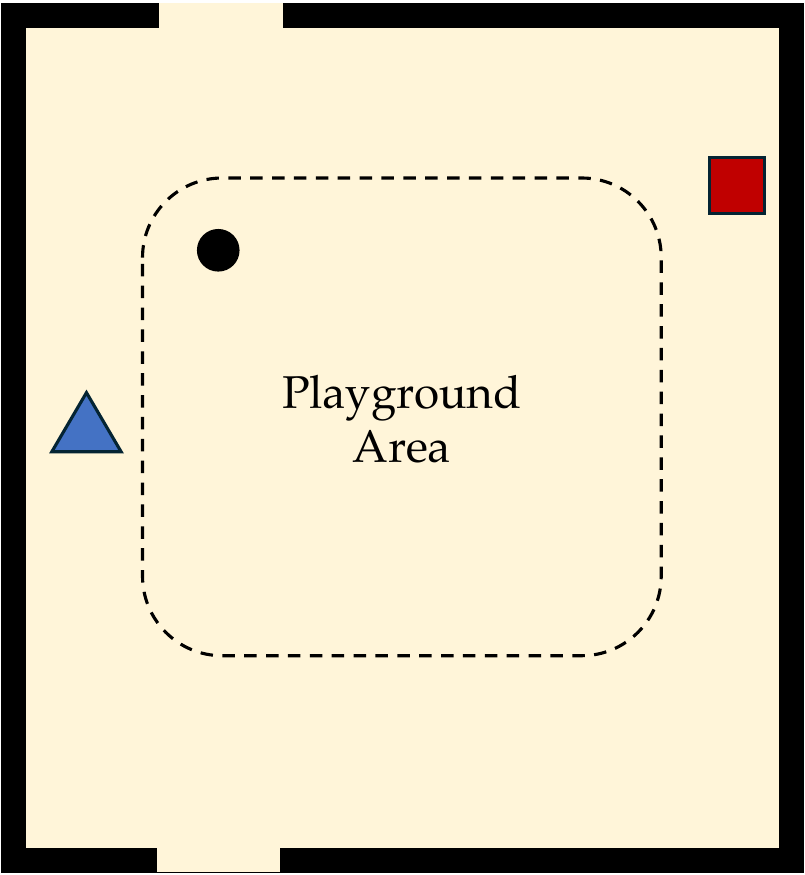}
        \caption{\textit{Activity recognition} room.}
        \label{fig:lab2}
    \end{subfigure}
    \caption{Layout of the rooms in which the experiments are performed. In the \textit{presence detection} case, the candidate enters/exits the room repeatedly, in the \textit{activity recognition} case, the candidate performs several activities in the marked area.}
    \label{fig:labs}
\end{figure}

In this work, we consider two separate \wifi sensing applications that are central in \gls{iot} forensics scenarios: \emph{presence detection} and \emph{activity recognition}.
In both cases, experiments involve collecting network traffic from consumer \gls{iot} devices to extract \gls{csi} measurements.
In the first use case (\emph{presence detection}), the transmitter and receiver are positioned on opposite sides of a room, with a candidate alternating between periods of being absent and moving within a designated area inside the room.
The dataset\footnote{Currently, the dataset is available upon request to the authors.} has been captured from the \wifi traffic of an \gls{iot} camera (transmitter) on a {20-MHz} channel in the {5-GHz} band; hence, every \gls{csi} feature vector consists of the amplitude of the 64 \gls{ofdm} subcarriers~\cite{palmese23collecting}.
The data was collected for an overall period of two hours using the Nexmon~CSI extractor~\cite{nexmoncsi2019} and balancing the different activities (present vs. absent).
We show in \cref{fig:lab1} the layout of the room.
The task is to determine the presence of a person in the room, i.e., a binary classification problem.

% \begin{figure}
%     \centering
%     \includegraphics[width=.5\columnwidth]{images/lab1b.pdf}
%     \caption{Room layout for the \textit{presence detection} scenario. A candidate enters/exits the room repeatedly.}
%     \label{fig:lab1}
% \end{figure}

In the second use case (\emph{activity recognition}), the experiments are performed in a different room where the transmitter and receiver exchange \wifi frames while the candidate acting as sensing target performs five different activities: walk, run, jump, sit, empty.
The dataset\footnote{The dataset is publicly available at \url{https://zenodo.org/records/7732595}.} has been collected from the \wifi frames of an 802.11ax access point injecting traffic at a constant rate of 150 frames per second for 80 seconds, using AX-CSI~\cite{gringoli2021axcsi} operating on a {160-MHz} channel in the {5-GHz} band.
As a result, each \gls{csi} feature vector consists of the amplitude of 2048 \gls{ofdm} subcarriers~\cite{cominelli23exposing}.
\Cref{fig:lab2} shows the layout of the room in which the activity recognition experiments are carried out.
In this case, the task is to determine the correct activity being performed among a set of possible activities.

% \begin{figure}
%     \centering
%     \includegraphics[width=.5\columnwidth]{images/lab2b.pdf}
%     \caption{Room layout for the \textit{activity recognition} scenario. A candidate performs different activities in the marked area.}
%     \label{fig:lab2}
% \end{figure}

\subsection{Presence Detection}
\label{ssec:presence-detection}

The \gls{csi} data is first pre-processed by filtering out non-informative subcarriers (e.g., guard bands), keeping only the \gls{csi} amplitude, and segmenting the data into 3-second groups.
For each group, the data is further divided into training and testing sets, each containing multiple windows of 64 frames, with a 32-frame overlap between consecutive windows and no overlap between training and testing subsets.
% We recall that our goal is to evaluate the impact of compression on \wifi sensing tasks, i.e., the tradeoff between sensing accuracy and compression ratio.

For the \emph{presence detection} task, we consider two different approaches: one based on a threshold classifier, trained on some features manually extracted from the data, and another one based on \gls{dl} techniques.

\begin{figure}[t]
	\centering
	\includegraphics[width=0.9\columnwidth]{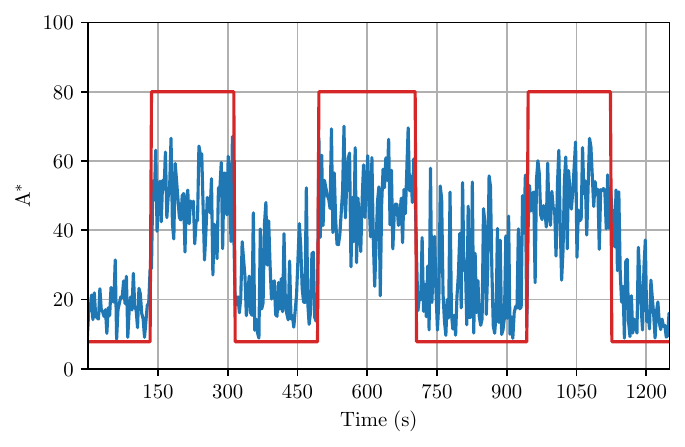}
	\caption{The value of $A^*$ over time (in blue). The red square wave indicates the ground truth when there is human presence (high value) or the room is empty (low value).}
	\label{fig:a_star_sample}
\end{figure}

For the threshold-based classifier, we consider the following techniques to compress the original data:
\begin{itemize}
    \item \textbf{SQ}:
    Classification is performed on the \gls{csi} amplitude where each subcarrier is quantized. The performance is expressed as a function of the quantization level (up to 8 bits per subcarrier).
    \item \textbf{VQ}:
    Classification is performed on the \gls{csi} amplitude where the entire \gls{csi} is vectorially quantized. The performance is expressed as a function of the quantization level (up to 8 bits per frame).
    \item \textbf{PCA+SQ}:
    first, we apply the \gls{pca} transform learned from training data and store each component as a 32-bit floating-point value. Then, classification is performed on the resulting frame where each component is quantized (up to 8 bits per component). We start with just one principal component and then consider more components.
    \item \textbf{PCA+VQ}:
    first, we apply the \gls{pca} transform learned from training data and store each component as a 32-bit floating-point value.
    Then, classification is performed on the vector quantization of the resulting frame (up to 8 bits per frame). We start with just one principal component and then consider more components.
\end{itemize}
Hence, the output of the Compression stage (cf. \cref{fig:methodology}) is a compressed version of the \gls{csi} amplitude dataset, containing time windows of \wifi frames.
We summarize the content of each time window using a single feature that quantifies the \textit{amount of variation} of the \gls{csi} data in that window.
Specifically, since each time window $k$ contains several \wifi frames, we first compute the sample standard deviation $\sigma^i_{k}$ of the \gls{csi} amplitude over each subcarrier $i$.
Then, we compute the average standard deviation over all the subcarriers in the time window $k$ to describe the window content through the single-valued feature:
\begin{equation*}
    A^*_{k} = \mathbb{E}\left\{\sigma_{k}^{i}\right\} \quad
\end{equation*}
which is used to feed a binary threshold classifier.
\Cref{fig:a_star_sample} shows the values of the feature $A^*$ when alternating user presence and absence, revealing that this feature is strongly related to the human presence in the environment, and thus validating our approach.

\begin{figure}[t]
	\centering
	\includegraphics[width=\columnwidth]{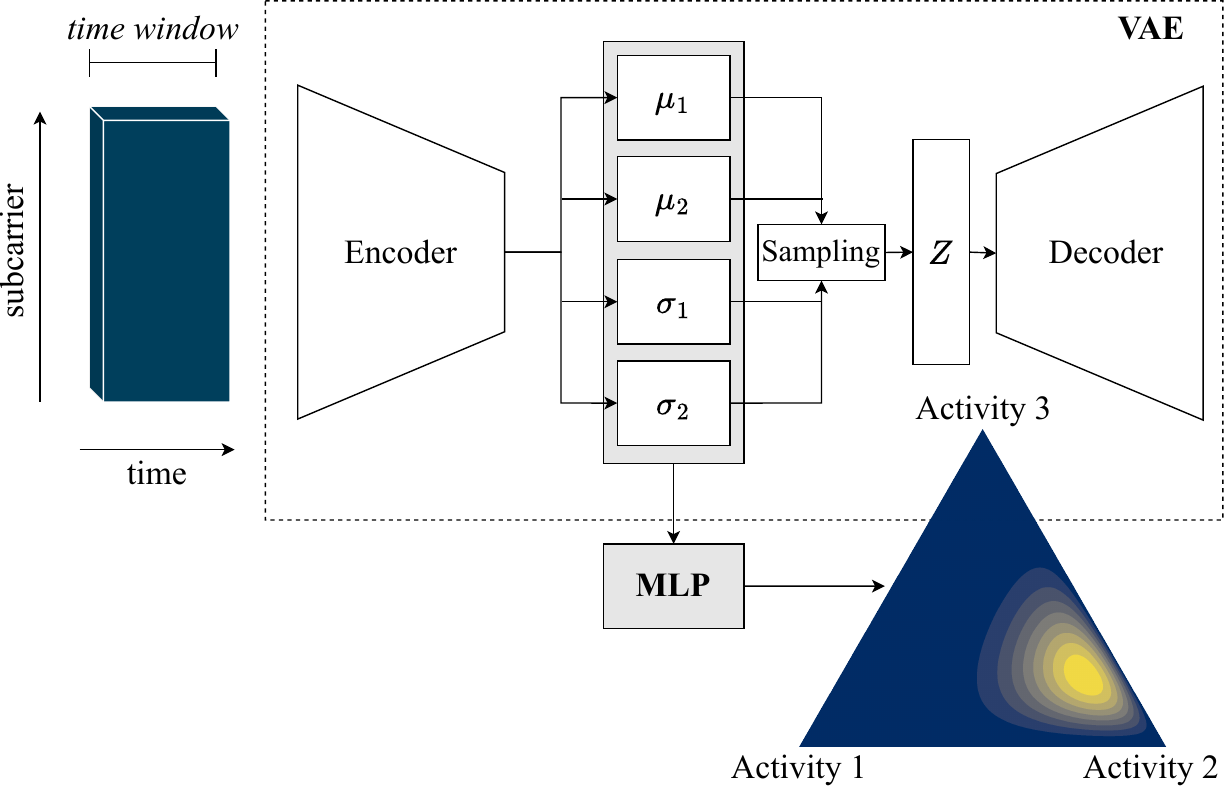}
	\caption{Sketch of the pipeline for the \acrshort{dl}-based classification approach when using the \acrshort{vae} to compress the dataset.}
	\label{fig:vae_arch}
\end{figure}

In the \gls{dl}-based approach, we replace the manual feature engineering with an automatic learning process.
Instead of relying on the threshold classifier, we feed the compressed dataset to a \gls{mlp} that has been trained to perform a binary classification of the current state of the room: presence or empty room.
In addition to the traditional compression schemes described before, we introduce in the compression stage (cf., \cref{fig:methodology}) a new technique based on a \gls{vae}.
The procedure based on the \gls{vae} and \gls{mlp} is sketched in \cref{fig:vae_arch}.
We highlight that the \gls{vae} could not be used with the manual feature engineering described before because the latent space of the \gls{vae} does not provide a convenient and explainable interpretation of its values (as opposed to $A^*$).
However, the latent space variables of the \gls{csi} can still be quantized with different levels of granularity, up to a maximum of 256 bits.
For this reason, we can define also in this case two possible compression techniques: \textbf{VAE\_SQ} and \textbf{VAE\_VQ}, based respectively on the scalar and vector quantization of the \gls{vae}'s latent space.

% In the \gls{dl}-based approach, we replaced manual feature extraction with an automated process based on a \gls{vae} as central component.
% In particular, instead of relying on a threshold algorithm on a single feature to determine whether a person is present in the room, a \gls{mlp} is utilized to perform a classification (of the presence or the target activity) based on the latent-space variables of a \gls{vae}.
% Specifically, the dataset is prepared following the same pre-processing steps as before, and training data is used to train the \gls{vae}. 
% In addition to the previously described experiments, this approach allows us to explore the scenario in which the latent space (i.e., the output of the VAE) was quantized at progressively higher levels of granularity, up to a maximum of 256 bits.
% From now on, these experiments will also be referred to as \textbf{VAE\_SQ} and \textbf{VAE\_VQ}.

For the \textit{presence detection} task, the training dataset has 56000 samples, while the test dataset has 14700 samples.
The average storage requirement for each original frame after pre-processing is 1792 bits.

\subsection{Activity Recognition}

In the \textit{activity recognition} dataset, 12500 \wifi frames have been collected for each activity (80 seconds of recording with a rate of 150~frames/s).
During the pre-processing step, the dataset is divided into training and testing subsets as follows.
First, every activity is split into 5 time windows of 16 seconds each.
Within each window, 9 seconds are assigned to the training set and 3 seconds to the testing set; the remaining seconds serve as a buffer to avoid overlap between the training and testing sets.
Then, we consider sequences of \gls{csi} data using a sliding window of 450 frames with a step size of one single sample to maximize data utilization during training and ensure sufficient data for both training and testing.

In the \textit{activity recognition} scenario, we utilize the same techniques as in the \gls{dl}-based approach for the \textit{presence detection} scenario.
The manual feature engineering is not considered because it is not powerful enough to ``capture'' the subtle variations of the \gls{csi} to discriminate among several different activities.
Using the same naming scheme introduced in \cref{ssec:presence-detection}, we evaluate the following compression techniques: \textbf{PCA+SQ}, \textbf{PCA+VQ}, \textbf{VAE\_SQ}, and \textbf{VAE\_VQ}.

In this scenario, the training dataset has 33750 samples, while the test dataset has 11250 samples.
Since the dataset has been collected using 160-MHz 802.11ax frames, the average storage requirement for each original frame after pre-processing is about 64~kb.

\section{Results}
\label{sec:results}

In this section, we analyze how lossy compression techniques affect the sensing performance across the two scenarios considered: \textit{presence detection} and \textit{activity recognition}.
Instead of focusing on the absolute sensing accuracy, we measure the performance loss in terms of the relative drop in F1-Score compared to the baseline using uncompressed \gls{csi} data, offering a metric for comparing the impact of different compression schemes.

\begin{figure*}
    \centering
    \begin{subfigure}[b]{0.49\linewidth}
        \includegraphics[width=\linewidth]{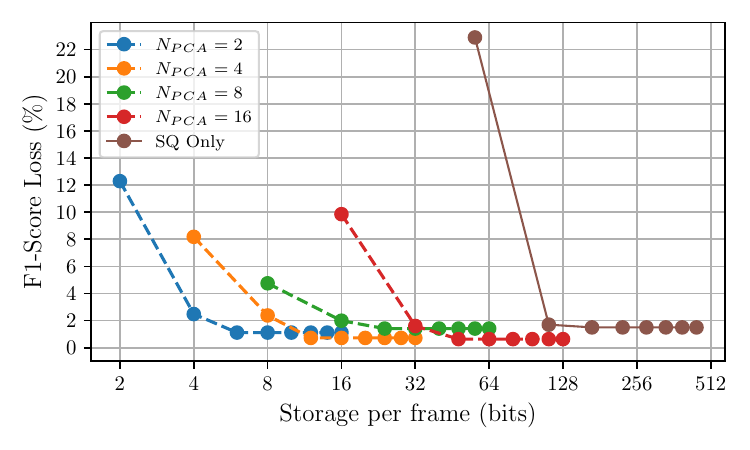}
        \caption{Relative F1-Score loss with scalar quantization.}
        \label{fig:pca_sq_noVAE}
    \end{subfigure}
    \hfill
    \begin{subfigure}[b]{0.49\linewidth}
        \includegraphics[width=\linewidth]{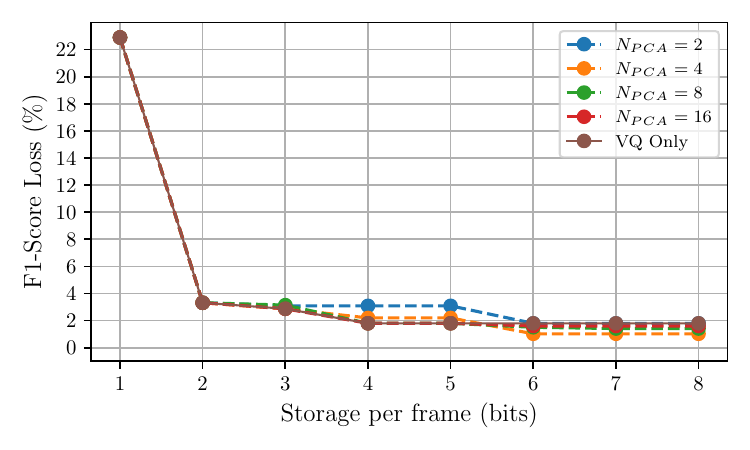}
        \caption{Relative F1-Score loss with vector quantization.}
        \label{fig:pca_vq_noVAE}
    \end{subfigure}
    \caption{Relation between compression rate (in bits per frame) and performance loss using the feature-based classifier in the \textit{presence detection} scenario. The uncompressed rate is 1792 bits per frame.}
    \label{fig:relative_f1_score_loss_FB}
\end{figure*}

\subsection{Presence Detection}
\label{ssec:presence-results}

\Cref{fig:relative_f1_score_loss_FB} summarizes the results for the \textit{presence detection scenario} using the feature-based classifier.
In the figures, we compare the classification performance and storage requirements of several PCA+SQ and PCA+VQ compression schemes.
The x-axis represents the average number of bits required to encode each \wifi frame under a given compression scheme, while the y-axis indicates the F1-Score loss relative to the original, uncompressed data (cf., \cref{fig:methodology}).
For each method, we vary the quantization resolution from 2 to 256 levels (i.e., from 1 to 8 bits per value) to assess how the compression ratio affects the performance.
In particular, the first thing we notice is that using more quantization bits minimizes the performance loss of the presence detection classifier.
While this observation may be trivial, it represents a sanity check on the outcomes of our experiments.

In the scalar quantization case (\cref{fig:pca_sq_noVAE}), the application of \gls{pca} significantly reduces both the storage required \textit{and} the performance loss with respect to just quantizing the original \gls{csi} values (SQ\_Only).
By retaining only the most informative linear combinations of the original subcarriers, \gls{pca} effectively reduces the \gls{csi} dimensionality without compromising the presence detection task.
Notably, using just two principal components---each quantized with only 3 bits, resulting in a total of 6 bits per frame---results in less than a 2\% loss in F1-Score.
Increasing the number of \gls{pca} components yields slightly improved accuracy but comes at the cost of an increase in storage.
These results suggest that aggressive dimensionality reduction via \gls{pca} is particularly effective when paired with scalar quantization.

In contrast, the benefits of \gls{pca} become far less pronounced when combined with vector quantization, as illustrated in \cref{fig:pca_vq_noVAE}. In this case, the application of \gls{pca} before VQ offers negligible improvements in the performance/compression trade-off.
This outcome can be attributed to the way vector quantization operates: the entire \gls{csi} vector is directly mapped to one of $2^L$ representative codewords, where $L$ is the number of bits used for indexing.
Since the quantization applies to the entire feature vector, reducing its dimensionality through PCA does not significantly alter the rate or structure of the compressed representation.
Interestingly, even without \gls{pca}, VQ achieves a favorable balance between performance and storage, with just 2 bits per frame already yielding satisfactory results (F1-Score loss below 4\%).
However, as the bitrate increases, the performance of SQ and VQ converges, with minimal differences beyond 8 bits per frame.

\begin{figure*}
    \centering
    \begin{subfigure}[b]{0.49\linewidth}
        \includegraphics[width=\linewidth]{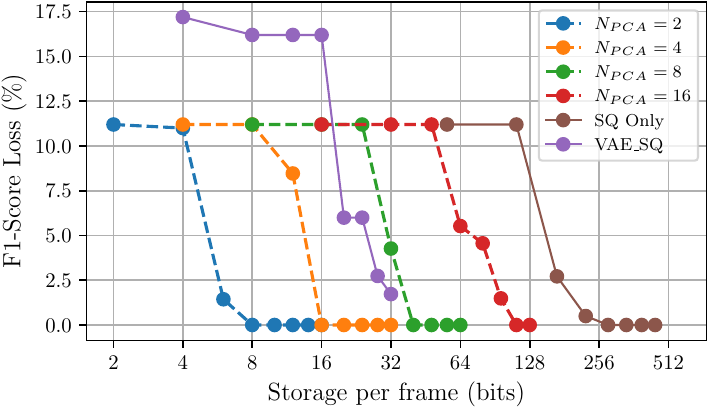}
        \caption{Relative F1-Score loss with scalar quantization.}
        \label{fig:pca_sq_VAE}
    \end{subfigure}
    \hfill
    \begin{subfigure}[b]{0.49\linewidth}
        \includegraphics[width=\linewidth]{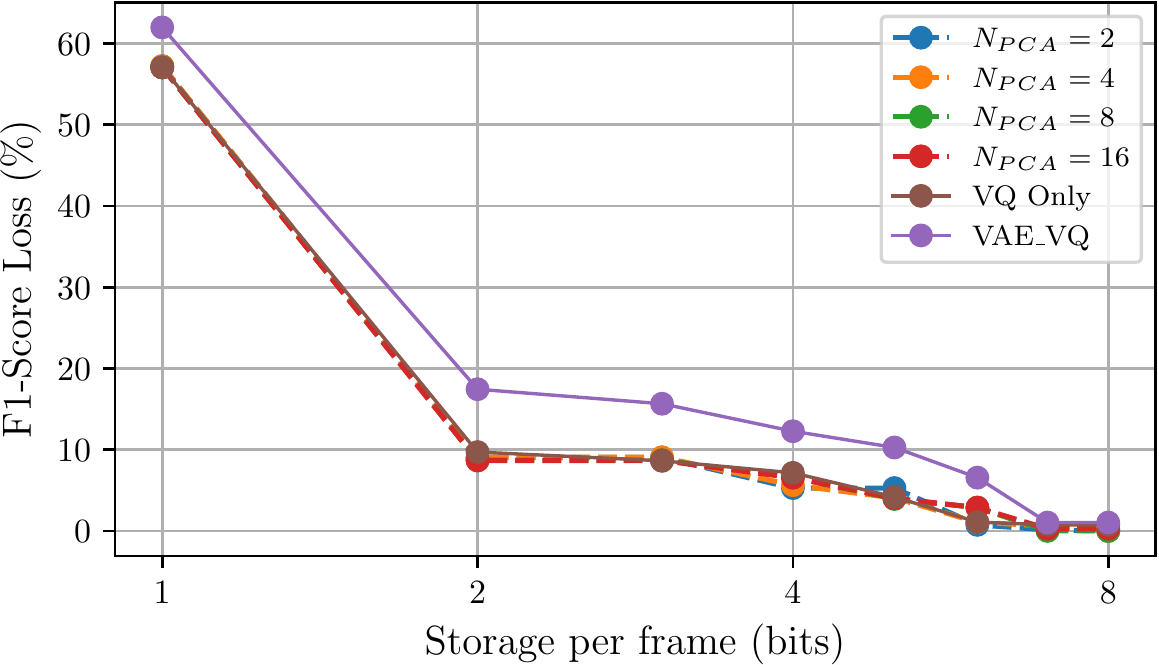}
        \caption{Relative F1-Score loss with vector quantization.}
        \label{fig:pca_vq_VAE}
    \end{subfigure}
    \caption{
    Relation between compression rate (in bits per frame) and performance loss using the \acrshort{dl} classifier in the \textit{presence detection} scenario.
    The uncompressed rate is 1792 bits per frame.}
    \label{fig:relative_f1_score_loss_VAE}
\end{figure*}

\Cref{fig:relative_f1_score_loss_VAE} shows the corresponding results using the \gls{dl} classifier. The trend observed for scalar quantization with \gls{pca} (\cref{fig:pca_sq_VAE}) remains consistent with the feature-based classifier.
When the first $N_{PCA}$ components are quantized using $L$ bits per value, the sensing performance improves steadily with the bitrate.
At higher bitrates, the F1-Score loss approaches zero, indicating that the \gls{mlp} classifier can effectively learn from compressed data, provided the quantization is sufficiently fine.
However, unlike the feature-based approach, the \gls{mlp} does not yield interpretable features and this may limit its utility and transparency in forensics applications.

The \gls{vae}-based compression scheme (VAE\_SQ) integrated with the MLP classifier achieves intermediate performance, falling between the results for $N_{PCA} = 4$ and $N_{PCA} = 8$.
This relatively modest performance may be due to the challenge of learning a well-separated latent space from the limited dataset used for training.
In other words, the \gls{vae} struggles to cluster the compressed features in a way that supports reliable classification.
Still, the underlying compression mechanism remains a viable option, to be validated with larger datasets.

The relative performance of \gls{pca} in VQ scenarios is shown in \cref{fig:pca_vq_VAE}.
As in the case of the feature-based classifier, the results confirm that \gls{pca} offers limited advantage in this context, and the number of retained \gls{pca} components has little effect when paired with vector quantization.
F1-Score losses in most cases range between 5\% and 15\%, with the best configurations sitting at around 2\% as the bitrate increases.
The VAE\_VQ architecture performs slightly worse for extreme compression rates but then converges to the PCA+VQ results at around 7 bits per frame.

It is important to highlight that compression methods optimized for inference accuracy may degrade the reconstructability of the original signal.
This is especially evident in the case of PCA+VQ with extreme compression ratios (e.g., almost 1800:1 if using 1 bit per frame), where the signal is effectively reduced to a binary indicator of presence or absence.
While suitable for task-specific sensing, such methods severely limit the ability to perform other types of forensic analysis based on signal reconstruction.
Taken together, these findings suggest that \gls{pca} combined with scalar quantization may offer the best sensing performance/compression tradeoff for the \textit{presence detection} task.
Indeed, our results show that it is possible to reduce data to just 8 bits per frame while preserving nearly the same sensing performance as with uncompressed \gls{csi}.

\subsection{Activity Recognition}

\begin{figure*}
    \centering
    \begin{subfigure}[b]{0.49\linewidth}
        \includegraphics[width=\linewidth]{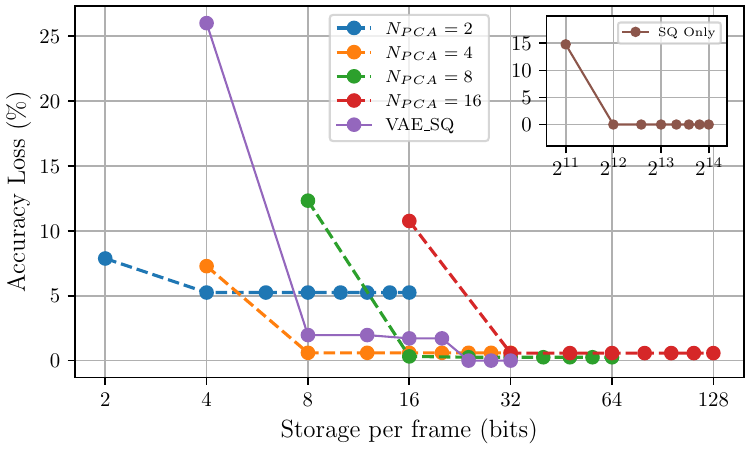}
        \caption{Relative F1-Score loss with scalar quantization.}
        \label{fig:pca_sq_AR}
    \end{subfigure}
    \hfill
    \begin{subfigure}[b]{0.49\linewidth}
        \includegraphics[width=\linewidth]{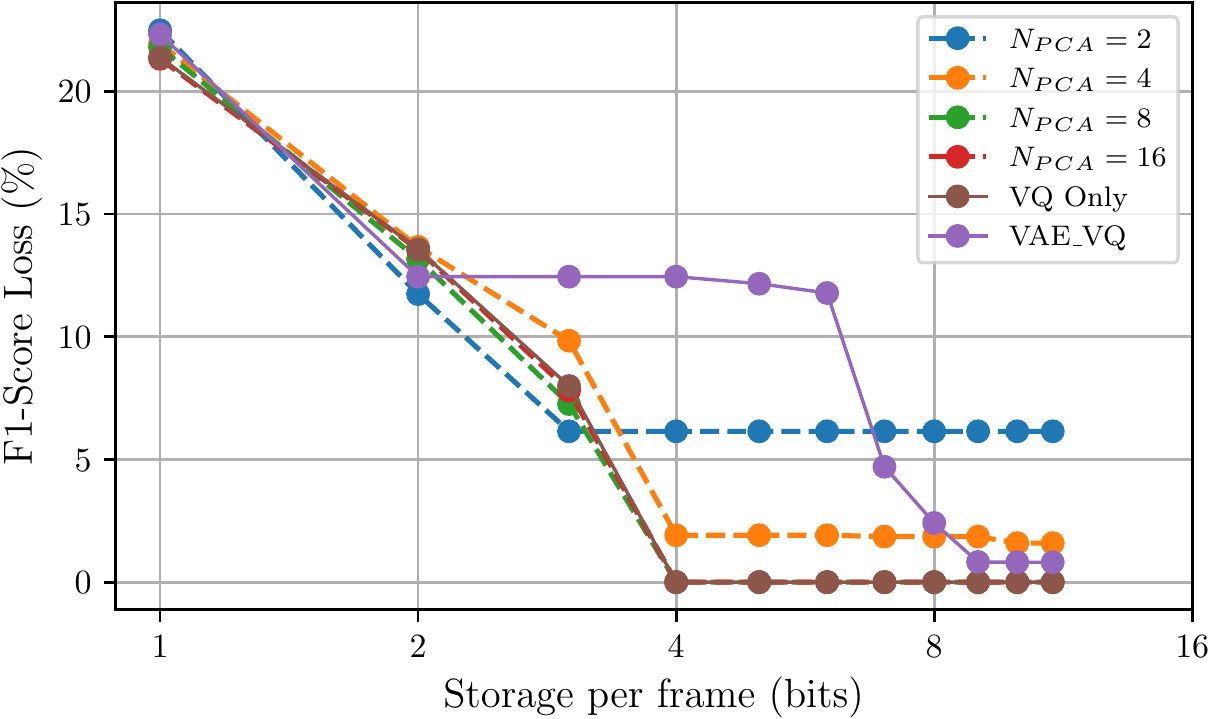}
        \caption{Relative F1-Score loss with vector quantization.}
        \label{fig:pca_vq_AR}
    \end{subfigure}
    \caption{
    Relation between compression rate (in bits per frame) and performance loss using the \acrshort{dl} classifier in the \textit{activity recognition} scenario.
    The uncompressed rate is about 64 kb per frame.}
    \label{fig:relative_acc_loss_VAE}
\end{figure*}

For the more complex \textit{activity recognition} scenario, we exclusively adopt the deep learning classifier, as the simple threshold-based classifier fails to adequately distinguish between nuanced variations in \gls{csi} data associated with different human activities.
The corresponding results for this task are presented in \cref{fig:relative_acc_loss_VAE}.
As before, \cref{fig:pca_sq_AR} highlights the benefit of performing PCA before scalar quantization.
This is particularly critical here due to the substantially larger number of subcarriers involved (2048 compared to 56 in the presence detection task).
Remarkably, quantizing just the first four principal components with 2 bits each (8 bits per frame) is sufficient to almost match the sensing performance of the uncompressed data.
The VAE-based compression again yields results that are comparable to PCA+SQ with $N_{PCA} = 4$ or $N_{PCA} = 8$, reflecting the same tradeoff observed in the presence detection task.
Although the VAE does not outperform PCA even in this setting, its potential for learned, task-specific representations shows opportunities for improvement with enhanced training or larger datasets.

In the case of vector quantization (\cref{fig:pca_vq_AR}), performance suffers significantly when only two PCA components are retained, achieving a minimum F1-Score loss of about 6\%.
This degradation can be attributed to the complexity of the activity recognition task, which requires more detailed and fine-grained features than presence detection.
Increasing the number of PCA components (e.g., $N_{PCA} \geq 8$) alleviates this problem and results in improved tradeoffs between compression and classification performance.
We highlight the fact that compressing 160-MHz 802.11ax \gls{csi} data to 4 bits per frame implies an impressive compression ratio of about 16000:1. 

Overall, the results reaffirm that PCA-based pre-processing is highly effective in scenarios where scalar quantization is employed.
In contrast, its utility diminishes in vector quantization settings, where the benefits of dimensionality reduction are largely negated by the encoding process.
Nonetheless, following the discussion at the end of \cref{ssec:presence-results}, PCA+SQ continues to provide the most favorable balance between compression and sensing accuracy across both tasks evaluated.

\section{Related Work}
\label{sec:related}

In this section, we briefly review the existing literature on (i) \gls{csi}-based human sensing and (ii) compression of \gls{csi} data, highlighting that while both areas have been explored independently, the problem of finding an optimal tradeoff between sensing accuracy and compression schemes in \gls{iot} forensics applications remains under-addressed.

\subsection{Human Sensing with Wi-Fi CSI}

Research on human sensing using \wifi signals has grown steadily since the introduction of the first \gls{csi} extraction tool for 802.11n frames~\cite{halperin2011tool} in 2011.
Today, surveys such as~\cite{ma2019survey,tan2022survey} provide a broad overview of sensing applications built on top of \gls{csi} data, which can be used for recognizing daily activities~\cite{wang2015understanding}, identifying gestures~\cite{li2016wifinger}, and detecting other types of human movement.
For instance, advanced analysis pipelines enable vital sign monitoring, including heart rate~\cite{liu2015tracking} and respiration~\cite{zhang2019breathtrack}.

Recent studies started to explore how low-cost devices like the ESP32 can be used to monitor the surrounding environment through \gls{csi}, enabling presence detection or movement direction estimation~\cite{hernandez2021adversarial}.
As commodity hardware becomes more capable of CSI extraction, ubiquitous \wifi sensing is likely to become a standard component of smart environments.
However, storing large volumes of high-dimensional \gls{csi} data for retrospective analysis---as required in forensics---poses challenges largely overlooked by the current literature.
This gap is precisely what motivated our investigation.

\subsection{Lossy Compression of CSI data}

On the other hand, research on compressing \gls{csi} data has largely focused on enhancing communication efficiency by reducing overhead in \gls{csi} feedback for beamforming and massive \gls{mimo} applications~\cite{guo2022overview}.
It has been shown that traditional transform coding approaches may sometimes outperform \gls{dl}-based methods in \gls{csi} compression~\cite{ornhag2023critical} due to the limited availability of datasets and the absence of standardized evaluation benchmarks.
Earlier efforts like~\cite{ferguson2010compression} achieved significant bitrate reductions using basic techniques, such as Huffman coding applied to quantized subcarrier amplitudes.
More recently, \cite{mismar2024adaptive} demonstrated the effectiveness of deep autoencoders for compressing \gls{csi} feedback in massive \gls{mimo} systems, suggesting that such architectures can support adaptive lossy compression based on channel dynamics, balancing reconstruction fidelity and compression ratio.

To our knowledge, \gls{csi} compression for the purpose of efficient \wifi sensing in \gls{iot} environments has seen far more limited attention.
One exception is~\cite{taso2018subcarrier}, which shows that removing selected \gls{ofdm} subcarriers can enhance localization accuracy in \gls{dl}-based systems.
Yet, this technique is primarily aimed at mitigating overfitting, not optimizing the storage/performance tradeoff.
Another notable exception is EfficientFi~\cite{yang2022efficientfi}, a \gls{dl} architecture built on a \gls{vae} that effectively reduces \gls{csi} dimensionality while preserving sensing accuracy.
Although promising, EfficientFi relies on substantial training data and large model sizes, making it less suitable for low-power \gls{iot} devices due to higher computational demands.
In contrast, in this work we explore also lightweight compression methods that better align with forensic requirements, achieving strong performance while improving result interpretability and resource efficiency.

\section{Conclusions}
\label{sec:conclusions}

In this work, we investigate the impact of lossy compression techniques on human sensing using the \wifi channel state information, studying the tradeoff between the achievable compression ratio and the sensing performance.
In our analysis, we consider both traditional and deep learning-based approaches.
In detail, we analyze the behavior of several compression techniques in two sensing scenarios: human presence detection and activity recognition.
These scenarios are particularly relevant in forensics applications because this information can be stored and retrieved as a source of evidence.

Experimental results show that combining \gls{pca} with simple quantization methods can already achieve a good balance between performance and storage efficiency.
While scalar quantization offers simplicity and ease of implementation, vector quantization consistently delivers superior compression with minimal penalty on performance.
In our experiments, deep learning-based compression methods have shown potential but did not match the simplicity and efficiency of the \gls{pca}, revealing the training phase effort does not bring clear advantages.
Overall, the results obtained in this work highlight how traditional compression techniques are still relevant in optimizing \wifi sensing systems, especially in low power \gls{iot} scenarios, achieving a compression ratio of up to 16000:1 without severely compromising the classification accuracy.

\section*{Acknowledgments}
This work was supported by the European Union and the Italian Ministry for University and Research (MUR) through the PRIN project ``COMPACT'' (Mission 4, Component 1, CUP D53D23001340006) and the Extended Partnership MICS (PE00000004, CUP D43C22003120001) under the Italian National Recovery and Resilience Plan (NRRP).
All projects are funded by NextGeneration EU.

\balance

\bibliographystyle{IEEEtran}
\bibliography{bibliography}

\end{document}